\documentclass[12pt]{article}
\usepackage{cite}
\usepackage{pifont}
\usepackage{times} 
\usepackage{amssymb,amsbsy,amsmath,amsthm}
\usepackage{xcolor}
\newcommand\revision[1]{{\color{black} #1}}
\newcommand\email[4]{#1@#2.#3.#4}
\newcommand\tr{\operatorname{Tr}\,}
\newcommand\spec{\operatorname{Spec}\,}
\def\C{{\mathbf{C}}}

\def\Z{{\mathbf{Z}}}
\def\N{{\mathbf{N}}}
\def\H{{\mathcal{H}}}

\def\eq#1{(\ref{#1})}

\renewcommand\hat{\widehat}

\def\viz{\emph{viz.}}
\newcommand\pa{\partial}

\theoremstyle{definition}
\newtheorem{example}{Example}
\title{Beta-gamma system, pure spinors and Hilbert series of arc spaces}
\author{
Chandrasekhar Bhamidipati\thanks{\email{chandrasekhar}{iitbbs}{ac}{in}} \\
\small School of Basic Sciences, 
\small Indian Institute of Technology Bhubaneswar \\
\small Bhubaneswar 751~007, India.
\and 
Koushik Ray \thanks{\email{koushik}{iacs}{res}{in}} \\
\small Department of Theoretical Physics,
\small  Indian Association for the Cultivation of Science \\
\small  Calcutta 700~032. India.
}
\date{}
\begin{document}
\setcounter{page}{1}
\maketitle
\thispagestyle{empty}
\begin{abstract}
\noindent 
Algorithms are presented for calculating the partition function of 
constrained beta-gamma systems in terms of the generating 
functions of the individual fields of the theory, the latter obtained 
as the Hilbert series of the arc space of the algebraic variety  
defined by the constraint. Examples of a beta-gamma system on a complex
surface with an $A_1$ singularity and pure spinors are worked out and
compared with existing results.
%
\end{abstract}
\section{Introduction}
A beta-gamma system is a two-dimensional conformal field theory 
modelled after the $b$-$c$ ghost system  
with a set of possibly bosonic complex
fields, denoted $\gamma$ and their canonical
conjugates, denoted $\beta$. It can be related to a certain large volume 
limit of the two-dimensional non-linear sigma-model with
the fields $\gamma$ identified as the complex coordinates of the target space
\cite{wit1,nek1,Tan:2006qt}. 
A beta-gamma system is said to be free, field theoretically, or flat,
geometrically, if the fields $\gamma$ and $\beta$ satisfy the commutation 
relations for free fields. In particular, a pair of $\gamma$'s commute.
The set of $\gamma$'s  then corresponds to  
the coordinates of the complex affine target space. A beta-gamma system
is said to be curved, if the target space is curved. In this article we shall
restrict to curved systems obtained as the coordinates of the target space
satisfy one or more algebraic equations. This is obviously equivalent to
imposing constraints on the $\gamma$'s. 

The action of a free beta-gamma system is linear in both the fields 
$\gamma$ and $\beta$. The partition function of this field theory is
obtained as the generating function of degeneracy of operators graded 
by quantum numbers associated to conserved charges of 
the classical action. The partition function of a constrained or 
curved system is then the generating function of degeneracy of 
operators satisfying the constraints.

The partition function can be computed by counting
operators possessing equal conserved charges
obtained through multiplication of $\beta$'s, $\gamma$'s and their 
derivatives with respect to the world-sheet coordinate.
A direct construction of operators, however, becomes
intractable in the presence of constraints,
save for the simplest of instances, as the derivatives of the
fields $\gamma$ and $\beta$ too are constrained by the derivatives of the
constraints to all orders. 
Partition function of curved beta-gamma systems in several instances
have been obtained by resorting to more indirect means 
\cite{gp,Grassi:2007va,aisaka,ours}. 
A special case, which serves as the motivation for the majority of studies of
the beta-gamma system in recent times, is
the pure spinor constraint which is a quadratic one arising in an attempt to 
write a super-Poincar\`e invariant world-sheet string theory \cite{berkovits}.
The partition function of pure spinors  has been
obtained as the character of representations of the $SO(8)$ group 
\cite{Berkovits:2005hy,morales,arroyo,aabn}.
In a variety of other examples the constraints are
not quadratic. In the case where the target space can be realized 
as an orbifold, for example,  
$\C^2/\Z_N$ or $\C^3/\Z_M\times\Z_N$, with integral $M$ and $N$, partition 
functions of beta-gamma systems have been 
obtained by lifting the geometric orbifold action to the partition function 
of the affine spaces $\C^2$ and $\C^3$, respectively \cite{ours}. 
This, however, relies upon the affine parametrization of the orbifolds.

In the present article we consider two examples of constraints. 
The first is a quadratic one among three $\gamma$'s, the other being
pure spinors, which also obeys a set of quadratic constraints. We use the
constraints directly without solving them, thereby avoiding any reference to
the affine parametrization.  
Regarding the constraints in $\gamma$'s as describing an 
algebraic variety embedded in the affine space of the 
unconstrained ones the contribution of the various modes of
$\gamma$'s to the partition function 
is given by the Hilbert series of the arc space of the variety.
However, in both the instances considered here the varieties possess an
isolated singular point. This renders the definition of the conjugate fields
non-unique. The total partition function is then obtained by 
resorting to some prescription. One efficient prescription is to implement
the so-called field-antifield symmetry of the partition function
in a multiplicative fashion
\cite{gp}. We show that it can also be obtained from the 
combination of various modes of the fields which are invariant under a
certain gauge transformation that keeps the action unchanged modulo the
constraints, provided the $\beta$'s are subjected to the same constraints. 
We exhibit the computations explicitly for two cases. We obtain the partition
function of a beta-gamma system on the rational double point surface 
singularity in both the ways and compare with the result obtained 
earlier \cite{ours} by realizing the target space as an orbifold. 
We find that the latter prescription fares slightly better when compared 
with the orbifold results. This computation uses the known description of 
resolution of surface singularities in terms of arc spaces. 
For the pure spinors this description is not known. We obtain the partition
function by implementing the field-antifield symmetry on the
contribution of the pure spinors obtained as the Hilbert series of the arc 
space of the pure spinor constraint. 
This is different from implementing the 
field-antifield symmetry at every order of mass separately.
Obtaining the Hilbert series 
entails a computation of Gr\"obner basis of the ideal generated by the pure
spinor constraint  by considering $10m$ 
equations in $16m$ variables
for every mass level $m$. The algorithm for this computation is rather 
simple and has been implemented in \verb|Macaulay2| \cite{m2}. 
The results match with the existing ones up to the first mass level. 

In section~\ref{sec:quad} we begin by recalling some features of 
the beta-gamma system and its partition function and lay out the two
prescriptions used to evaluate the partition function. 
In section~\ref{sec:arc} we recall the notion of arc spaces 
and the blow up of surface singularities in these terms.
We use both prescriptions to compute the partition function of the beta-gamma
system on the rational double point surface singularity in the following
section, comparing the results. 
In section~\ref{sec:ps} we obtain the partition function of the pure spinor
system up to the first mass level by implementing the field-antifield
symmetry. We conclude in section~\ref{sec:concl}. 
\section{$\beta$-$\gamma$ system on $\C^d$}
\label{sec:quad}
\subsection{Flat system}
A beta-gamma system on the $d$-dimensional complex affine space $\C^d$ is
a two-dimensional conformal field theory of a set of 
complex fields $\{\gamma^i\}$ of vanishing conformal dimension and their 
canonical conjugates $\{\beta_i\}$, $i=1,2,\cdots, d$. 
On the two-dimensional space, henceforth referred to as the world-sheet,
the conjugate fields are one forms, namely, $\beta_i=\beta_{iz}\
dz+\beta_{i\bar{z}}\ d\bar{z}$, where $z$ designates the coordinate of the
world-sheet and a bar denotes its complex conjugate. 
For a flat beta-gamma system the fields $\gamma$ are identified with 
the coordinates of the coordinate
ring of the target space $\C^d=\C[x_1,x_2,\cdots, x_d]$ as $\gamma^i=x_i$.
The coordinates commute pairwise as do the conjugates thereby having
trivial operator products. The operator product between a $\beta$ and a
$\gamma$, on the other hand, is taken to be the free one, namely
\begin{equation} 
\gamma^i(z) \beta_j(z') \sim \delta^i_j\frac{dz'}{z-z'}.
\end{equation} 
The action for a beta-gamma system is written as 
\begin{equation}
\label{bgact}
S = \frac{1}{2\pi}\sum\limits_{i=1}^d\int \beta_i\bar{\pa}\gamma^i
\end{equation}
in the conformal gauge, 
where $\pa = \frac{\pa}{\pa z}$. 
The theory possesses two conserved currents, namely, 
the energy momentum tensor
and a $U(1)$ current corresponding to the scaling of the fields, 
\begin{equation}
\label{eq:scaling}
\gamma^i\longrightarrow \Lambda_i\gamma^i, \quad
\beta_i\longrightarrow\Lambda_i^{-1}\beta_i.
\end{equation}
The respective charges, namely,
$L_0=\oint dzz \beta_{iz}\partial\gamma^i$ and
$J_0 = \oint dz \beta_i\gamma^i$,
characterize the field theory.  Introducing the modular parameter
$q$ and another one, $t$,
corresponding to the scaling the partition function of
the beta-gamma system is written as the character
\begin{equation}
\mathcal{Z} = \tr (q^{L_0} t^{J_0}),
\end{equation}
where $\tr$ signifies a trace with respect to the states of the
Hilbert space of the theory.

Assuming that the fields possess mode expansions
\begin{equation}
\label{hwmode}
\begin{split}
\beta_i(z)&=z^{-1/2}\sum_{n\in\Z} z^{-n-1}\beta_{i(n+1)}\\
\gamma^i(z)&=z^{1/2}\sum_{n\in\Z} z^{-n-1}\gamma^i_{(n)}
\end{split} 
\end{equation} 
and the existence of a vacuum $|0\rangle$
to obey the highest weight conditions 
\begin{equation}
\label{hwrep}
\beta_{i(n+1)}|0\rangle =0\quad 
\gamma^i_{(n)}|0\rangle =0,\quad n\geqslant 0,
\end{equation} 
the character of the beta-gamma system on $\C^d$ is obtained as 
\cite{Guruswamy,Lesage}
\begin{equation}
\label{zcd}
\mathcal{Z}_{\C^d} = (\mathcal{Z}_{\C})^d,
\end{equation}
where the character of the affine complex plane is defined to be
\begin{equation}
\label{zc}
\mathcal{Z}_{\C} = \frac{1}{1-t}\prod_{n=1}^{\infty}
\frac{1}{(1-q^nt)(1-q^n/t)}.
\end{equation}
This can be interpreted as the generating function of degeneracy of monomials
of a given degree for $q$ and $t$, where each $\gamma$
contributes
a $t$, each $\beta$ contributes $q/t$ to the partition function while each
derivative $\pa$ contributes a $q$ \cite{ours}.
\subsection{Curved system}
\label{sec:curv}
One type  of curved beta-gamma systems is obtained from a flat one by imposing
constraints on the fields $\gamma$. The constraints we consider are
non-linear but algebraic, which are 
well-defined as the $\gamma$'s commute between themselves. 
This makes the target space into a non-affine algebraic variety. 
The various operators then correspond one-to-one with 
the regular functions formed from monomials on this variety. 
We present a means to evaluate the partition function as the Hilbert series of
the arc space of the variety.

More specifically, we deal with quadratic constraints given by
\begin{equation}
\label{constr:general}
\sum_{j=1}^{d}\Omega_{ij}\gamma^i\gamma^j=0,
\end{equation} 
with one or more  
constant $d\times d$ matrices $\Omega$. 
Computation of the partition function then entails
enumeration of monomials in the fields $\beta$ and $\gamma$ as well as
their derivatives with respect to the world-sheet coordinate modulo the
constraint and its derivatives to all orders.
The multiplication of fields in forming monomials are to be normal ordered as
usual, but this does not affect their number.

\revision{
Constructing monomials, alias operators,
involving $\beta$'s is ambiguous as they are not constrained \emph{prima
facie} unlike the $\gamma$'s.
They are to be constrained by prescribing extra  conditions.
Treating them as canonical momenta conjugate to the $\gamma$'s, as is in
fact necessitated by its connection with the sigma model \cite{nek1}
we end up with the usual problem of defining momenta on a singular space.
Indirect means of constraining the $\beta$'s are therefore to be devised.
We consider two different ways to evaluate the partition function of 
this theory, starting
from the separate contributions of the $\gamma$'s and $\beta$'s.
In both the methods the contribution from the $\gamma$'s alone , denoted
$Z_{\gamma}$, is obtained first by counting monomials constructed solely
from the $\gamma$'s, wherein the constraint \eq{constr:general} is taken
into account. The share from the $\beta$'s is then derived from $Z_{\gamma}$
using  symmetries of the action relating the $\gamma$'s and the $\beta$'s.

In the first method, implementing the so-called field-antifield symmetry
$Z_{\gamma}$ is split into two factors. The first 
is independent of $q$ arising from the contribution of the zero
modes. The other is a function of both $q$ and $t$ from 
the contribution of the massive modes. Thus,
\begin{equation}
\label{gsplit}
Z_{\gamma}=Z_0(t)Z_{m}(q,t), 
\end{equation} 
where the subscripts $0$ and $m$ refer to the zero and non-zero mass modes. 
The total partition function is obtained as \cite{gp}
\begin{equation}
\label{pf:g}
Z(q,t)=Z_0(t)Z_{m}(q,t) Z_{m}(q,1/t). 
\end{equation} 
This method has been used previously in a ghost-for-ghost scheme for pure
spinors \cite{aabn}.

We propose an alternative using  the gauge invariance of the action 
\eq{bgact} under the transformation 
\begin{equation}
\label{gagtr:general}
\begin{split}
\delta\gamma^i=0,\qquad
\delta\beta_i = \sum_{i,j=1}^d\Omega_{ij}\gamma^j.
\end{split}
\end{equation} 
The action \eq{bgact} is invariant under this transformation
modulo \eq{constr:general}.
This imposes a restriction on the possible combinations of suitably defined
$\beta$'s due to the constraints on $\gamma$'s. Only 
$\beta$'s appearing in gauge invariant combinations are then counted in the
partition function. Two such combinations
at first and second mass levels, for example,
are the $U(1)$ current and energy momentum tensor, 
respectively, which are composite operators.
Although the number of gauge invariant operators at each mass level is 
finite, new operators emerge at each higher mass level, rendering the 
counting of such states intractable. As a consequence, the partition function
generically contains negative terms in $1/t$. 
This is a hurdle in obtaining the 
partition function of $\beta$-$\gamma$ systems in a closed form.

We use a new method to implement gauge invariance in the 
$\beta$-$\gamma$ system directly at the level of the partition function, 
thereby, giving a rationale for the omission of negative terms in $1/t$.  
Assuming the existence of the conjugate fields we obtain their separate
contribution to the partition function
by subjecting them to the same constraint as the $\gamma$'s, namely, 
\begin{equation}
\label{constr:beta}
\sum_{j=1}^{d}\Omega_{ij}\beta_i\beta_j=0.
\end{equation} 

Supposing we have a way of finding $Z_{\gamma}$, we can use the same
method to count the totality of monomials of $\beta$'s alone modulo this
constraint. Let us denote it by $Z_{\beta}$. 
\revision{ The full partition function 
of the theory is then obtained by combining $Z_{\gamma}$ and $Z_{\beta}$
in such a way that the condition of gauge invariance in \eq{gagtr:general}
is respected.

The na\"ive product $Z_{\gamma}Z_{\beta}$ actually counts the set of all possible monomials
constructed out of the fields $\beta$ and $\gamma$ and their derivative, 
satisfying, the constraints \eq{constr:beta} and \eq{constr:general} and their derivatives.
Out of this set, we need to subtract the gauge-noninvariant monomials, namely, the 
monomials which do not vanish modulo \eq{constr:general}. We implement this as follows.

We subtract $Z_{\beta}-1$. $Z_{\beta}$ is subtracted because the constraint \eq{constr:general}  is quadratic and the gauge transformation \eq{gagtr:general} of $\beta$ produces a single
power of $\gamma$. Thus, one checks that monomials constructed solely from $\beta$'s are not
gauge invariant. The unity is subtracted so as to avoid over counting the 
constant monomial twice. The partition function is thus}
\begin{equation} 
\begin{split}
Z'(q,t)&= Z_{\gamma}Z_{\beta} -Z_{\beta} +1 - Z_{\gamma} + Z_{\gamma} \\
&=Z_{\gamma}+ (Z_{\gamma}-1)(Z_{\beta}-1).
\end{split}
\end{equation}
There are more monomials to discard. Let us recall that $Z_{\gamma}$ is the
partition function of monomials in $\gamma$ which \emph{do not} vanish modulo
\eq{constr:general}. \revision{However, the 
constant monomial (the monomial $(\gamma_i)^0$ or unity) as well as
ones with single powers of $\gamma$'s, which are counted in $Z_{\gamma}$ as the
terms constant and linear in $t$, respectively,
can not arise from a combination of  $\beta$'s and $\gamma$'s by
\eq{gagtr:general}. These constitute non-invariant monomials. The totality of non-invariant monomials involving
both types of fields is then}
\begin{equation} 
\label{gau}
Z_{\beta}\left(Z_{\gamma}-1-t 
\left[\frac{dZ_{\gamma}(q,t)}{dt}\right]_{t=0}\right).
\end{equation}
Moreover, the  gauge transformation \eq{gagtr:general} on any monomial
converts a $\beta$ into a linear combination of $\gamma$'s, thereby changing
the grade of a monomial by a factor of $q/t^2$. \revision{ Thus, the above expression \eq{gau} is
to be subtracted from $Z'(q,t)$ after compensating for this change in grade. The resulting partition
function is}
\begin{equation}
\label{pf:formula}
Z(q,t)=Z_{\gamma}+(Z_{\beta}-1)(Z_{\gamma}-1)
-\frac{q}{t^2}\
Z_{\beta}\left(Z_{\gamma}-1-t 
\left[\frac{dZ_{\gamma}(q,t)}{dt}\right]_{t=0}\right).
\end{equation}

}

Let us recall that
to evaluate the partition function in either way, we need to find
$Z_{\gamma}$. For the second method we also need $Z_{\beta}$. These are
obtained as Hilbert series of the arc spaces of the varieties described by
\eq{constr:general} and \eq{constr:beta}, respectively, to which we turn
next.

\section{Arc spaces and Hilbert Series}
\label{sec:arc}
In this section we recall some features of the arc space of an
algebraic variety \cite{nash} and define its Hilbert series. 
Relation between Hilbert series of arc spaces in a single variable 
and partitions has been noted earlier \cite{rr:arc}. 
We restrict attention to complex numbers only but generalize the definition
of Hilbert series to graded rings to incorporate the grades of $q$ 
and $t$ pertinent to beta-gamma systems. Let $\C[[\xi]]$ denote the formal power series (Puiseux series) ring of polynomials
in a single variable $\xi$ over the field of complex numbers $\C$. In the
simplest case the arc
space of an algebraic variety defined by a 
polynomial equation $f=0$ in the coordinate ring $\C[x_1,x_2,\cdots, x_n]$ 
is the set of power series solutions to the equation $f(x(\xi))=0$, where
$x(\xi) = (x_1(\xi),x_2(\xi),\cdots, x_n(\xi)) \in \C[[\xi]]^n$, with 
each component a power series in the formal variable $\xi$. This generalizes
to more polynomials than one. 

More formally, let $\mathcal{M} = \spec\big(\C[x_1,x_2,\cdots, x_n]/(f_1,f_2,\cdots,
f_m)\big)$ be an algebraic variety defined by $m$ equations in the coordinate 
ring of $\C^n$. Let us write the coordinates $x_i$ as formal power series in a
formal variable $\xi$ as
\begin{equation}
\label{pui}
x_i = \sum\limits_{j=1}^{r}x_i^{(j)} \xi^j.
\end{equation} 
Substituting these series in the polynomials $f_k$, $k=1,2,\cdots, m$ and
truncating at the order $\xi^r$ we obtain the set of polynomials 
$F_k^{(l)} $ as the coefficient of $\xi^l$ in the expansion of
$f_k (x_1,x_2,\cdots, x_n)$. The $r$-th \emph{jet scheme} $M_r$ of $M$ is then
defined as
\begin{equation}
\mathcal{M}_r = \spec\left(
\frac{\C[x_i^{(j)}; 1\leqslant i\leqslant n, 0\leqslant j\leqslant r]}
{\{F_k^{(l)}; 1\leqslant k\leqslant m, 0\leqslant l\leqslant r\}}
\right) 
\end{equation} 
In particular, $\mathcal{M}_0=\mathcal{M}$, the variety itself and 
$\mathcal{M}_1 = T\mathcal{M}$, the tangent space. The \emph{arc space}
of $M$ is then defined as
\begin{equation}
\mathcal{M}_{\infty} = \spec\left(
\frac{\C[x_i^{(j)}; 1\leqslant i\leqslant n, j\in\N]}
{\{F_k^{(l)}; 1\leqslant k\leqslant m, l\in\N\}}
\right),
\end{equation} 
where $\N$ denotes the set of natural numbers, $\N = \{0,1,2,3,\cdots\}$.
In more mundane terms, the arc space on $\mathcal{M}$ 
is defined by the infinite set
of equations obtained at each order of $\xi$ by substituting an infinite
series of the form \eq{pui} into the defining equations $f_k=0$ of
$\mathcal{M}$ for $k=1,2,\cdots m$.

A generating function for monomials in the variables
$x_i^{(j)}$, modulo the relations $F_k^{(l)}$ is obtained
by bestowing a grade to the
variables $x_i^{(j)}$. This is defined to be the Hilbert series
of the  arc space $\mathcal{M}_{\infty}$, denoted
$\H_{f_1,f_2,\cdots, f_m}$ or $\H_{\mathcal{M}}$. 
Evaluation of the Hilbert series requires computation
of Gr\"obner basis from $F_k^{(l)}$, in general. However, for simple cases
this complication may not exist. We shall associate a grade $q^j t$ to the
variable $x_i^{(j)}$. The symbols are chosen to make the connection with the
beta-gamma system conspicuous. 
\subsection{Arc space of the singular quadric in $\C^3$}
Let us illustrate the computation of the Hilbert space with two 
simple examples. Further examples with a singly graded variable
exist in literature \cite{rr:arc}.
The arc space of the affine space $\C[x]$ consists of all the powers of
$x^{(j)}$ for all $j=0,1,\cdots,\infty$. 
The monomials are thereby obtained by arranging each of $x^{(j)}$ in
a geometric series $1+x^{(j)}+(x^{(j)})^2+(x^{(j)})^2 \cdots
= 1/(1-x^{(j)})$ and multiplying them as
$1/\prod_{j=1}^{\infty}(1-x^{(j)})$.  With the assignment of grades $q^jt$ to
$x^{(j)}$ as above then yields the Hilbert series of the arc space of $\C[x]$
as
\begin{equation}
\label{HS:C}
\mathcal{\H}_{\C} = \frac{1}{1-t}\prod_{n=1}^{\infty}\frac{1}{(1-q^nt)}.
\end{equation} 
Next let us work out the Hilbert series of the variety defined in
$\C[x_1,x_2,x_3]$ by the quadratic polynomial $f=x_1x_2-x_3^2$, corresponding
to the rational double point singular variety
\begin{equation} 
\label{ratdpt}
x_1x_2-x_3^2=0.
\end{equation}
This will provide part of the partition function $Z_{\gamma}$ of the 
beta-gamma system discussed in the previous section. 
Substituting the power series \eq{pui} in $x_1x_2-x_3^2$ we obtain the
polynomials 
\begin{gather}
F^{(0)}=x_1^{(0)} x_2^{(0)}-(x_3^{(0)})^2,\\
F^{(1)}=x_1^{(0)} x_2^{(1)}+ x_1^{(1)} x_2^{(0)} -2 x_3^{(0)}x_3^{(1)},\\
F^{(2)}=x_1^{(2)} x_2^{(0)}+ x_1^{(1)} x_2^{(1)}+x_1^{(0)} x_2^{(2)}
- (x_3^{(1)})^2-2 x_3^{(0)}x_3^{(1)},\\
\vdots
\nonumber
\end{gather}
at different orders of $\xi$.
According to our assignment of grades to the variables every $F^{(l)}$, being
quadratic, has $t$-grade $t^2$ and $q$-grade $q^l$. Now, considering only the
three relations on the nine variables $x_1^{(0)},\cdots, x_3^{(2)}$, the
Hilbert series is \cite{stanley,ours}
\begin{equation}
\frac{(1-t^2)(1-qt^2)(1-q^2t^2)}{(1-t)^3(1-qt)^3(1-q^2t)^3}.
\end{equation}  
Continuing \emph{ad infinitum} for the countably infinite quadratic equations
for the countable set of variables, we obtain the Hilbert series of the arc
space of the variety $f=0$ to be

\revision{
\begin{eqnarray} 
\label{hilb2:prod}
H_{x_1x_2-x_3^2}(q,t) &=&\frac{(1-t^2)}{(1-t)^3}
\prod\limits_{n=1}^{\infty}\frac{(1-q^nt^2)}{(1-q^nt)^3} \\
&=& \left(1+3 t+5 t^2+7 t^3+9 t^4+11 t^5+13 t^6+15 t^7+17 t^8+\cdots\right) \nonumber \\ 
&+& q \left(3 t+8 t^2+12 t^3+16 t^4+20 t^5+24 t^6+28 t^7+32  t^8+\cdots\right) \nonumber  \\
&+& q^2 \left(3 t+14 t^2+27 t^3+37 t^4+47 t^5+57 t^6+67 t^7+77 t^8+\cdots\right)  \nonumber \\
&+& q^3 \left(3 t+17 t^2+43 t^3+68 t^4+88 t^5+108 t^6+128 t^7+148 t^8+\cdots\right) \nonumber  \\
&+& q^4 \left(3 t+23 t^2+66 t^3+119 t^4+166 t^5+206 t^6+246 t^7+286 t^8+\cdots\right)  \nonumber \\
&+& q^5 \left(3 t+26 t^2+90 t^3+180 t^4+271 t^5+352 t^6+424 t^7+49  t^8+\cdots\right) \nonumber \\
\label{ser:gamma}
&+& \mathcal{O}\left(q^6\right) 
\end{eqnarray} 
Similarly, assuming that the $\beta$'s obey the same constraint and 
noting that they
do not possess zero modes, the Hilbert series for them is obtained as
\begin{eqnarray} 
\label{hilb2:beta}
H_{\beta_1\beta_2-\beta_3^2}(q,t) &=&
\prod\limits_{n=1}^{\infty}\frac{(1-q^{n+1}/t^2)}{(1-q^n/t)^3}  \\ 
& =& q\left(1+ \frac{3}{t}+\cdots\right)  +  
q^2 \left(\frac{5}{t^2}+\frac{3}{t}+\cdots\right) \nonumber \\ 
&+& q^3\left(\frac{7}{t^3}+\frac{8}{t^2}+\frac{3}{t}+\cdots\right) 
+ q^4\left(\frac{9}{t^4}+\frac{12}{t^3}+\frac{14}{t^2}+\frac{3}{t}+\cdots\right)\nonumber \\ 
&+&
q^5\left(\frac{11}{t^5}+\frac{16}{t^4}+\frac{27}{t^3}+\frac{17}{t^2}+\frac{3}{t}+\right)
+\mathcal{O}\left(q^6\right) 
\label{ser:beta}
\end{eqnarray} 
where the grade of $\beta_i^{(j)}$ is chosen to be $q^i/t$ for $i=1,2,3,\cdots$. }
\subsubsection{Contribution from blow up}
Resolution of rational surface singularities can be treated using arcs. 
The surface \eq{ratdpt} with an $A_1$ singularity 
at the origin is blown up with a $\mathbf{P}^1$. The single 
exceptional divisor corresponds to a truncation of the Puiseux series
\eq{pui} to \cite{nash,reguera}
\begin{equation}
\label{pui:tr}
x_i = x_i^{(1)} \xi,
\end{equation} 
for $i=1,2,3$.
Putting this truncated series in the constraint \eq{ratdpt} leads to the
single equations
\begin{equation}
x_1^{(1)}x_2^{(1)} - (x_3^{(1)})^2=0.
\end{equation} 
The Hilbert series is
\begin{equation}
\label{hbl}
\H_{Bl_0(x_1x_2-x_3^2)}(q,t) = \frac{1-q^2 t^2}{(1-qt)^3}. 
\end{equation} 
\section{Beta-Gamma system on surface with a rational double point} 
In this section we obtain the partition function of a beta-gamma 
system on the surface with an $A_1$ singularity
using the Hilbert series obtained above in
two different ways \eq{pf:g} and \eq{pf:formula} as discussed before. 
Let us note that the coefficients $x^{(j)}_i$ are in one-to-one
correspondence with the derivatives $\pa^j\gamma^{i}$ of the fields as well
as with the modes in \eq{hwmode}. The former identification is more better
suited for our purposes here. This allows the identification of the
Hilbert series as the relevant part of the partition function through
counting monomials in the fields and their derivatives. 
In the case of the affine space, there is no constraint.
Identifying the coefficients $x_i^{(j)}$ in \eq{pui} with 
$\pa^j\gamma^{i}$, $i=1,2,3$, the Hilbert series for each $\C$  
is given by \eq{HS:C}. Each component
of the arc space will have a conjugate with an inverse $t$-charge 
corresponding to the unconstrained $\beta$'s as well. Thus the
total partition function of a flat beta-gamma system is obtained by
augmenting $\H_{\C}$ in \eq{HS:C} with the contribution from the
conjugates, resulting into \eq{zc}. This can also be thought as an instance
of implementing the field-antifield symmetry according to \eq{pf:g}.

Let us now discuss the case of the quadratic constraint 
\eq{constr:general} with 
\begin{equation}
\Omega = \begin{pmatrix}
0 & 1&0\\
1&0&0\\0&0&-2
\end{pmatrix}.
\end{equation} 
The constraint is \eq{ratdpt} with the identification $x_i=\gamma^i$. 
This singular variety can also be looked upon 
as the orbifold $\C^2/\Z_2$, where the $\Z_2$ acts on
$(u,v)\in\C^2$ by changing signs of both. The partition
function has been computed earlier and written in a closed form \cite{ours},
by directly implementing the orbifolding on the partition function of the
affine space \eq{zcd} with $d=2$. 
The partition function \cite[eq. (29)]{ours} 
expanded in a series with respect to $q$ and $t$ is
\begin{equation}
\label{zorb}
\begin{split}
Z_{\C^2/\Z_2}(q,t) &=
(1+3 t^2+5 t^4+7 t^6+9 t^8+11 t^{10}+\cdots)\\
&\;\; +q
(4+12 t^2+20 t^4+28 t^6+36 t^8+44 t^{10}+\cdots)\\
&\;\; +q^2 (\frac{3}{t^2}+17+42
   t^2+70 t^4+98 t^6+126 t^8+154 t^{10}+\cdots)\\
&\;\; +q^3
(\frac{12}{t^2}+52+120 t^2+200 t^4+280 t^6+360t^8+440t^{10}+\cdots)\\
&\;\; +q^4
(\frac{5}{t^4}+\frac{42}{t^2}+147+320 t^2+525 t^4+735 t^6+945
   t^8+1155 t^{10}+\cdots)\\
&\;\; +q^5
(\frac{20}{t^4}+\frac{120}{t^2}+372+776 t^2+1260 t^4\\
& \qquad \qquad \qquad \qquad \qquad+ 1764t^6+2268t^8
+2772t^{10}+\cdots)+\mathcal{O}\left(q^6\right).
\end{split} 
\end{equation} 
The orbifold description and \eq{constr:general} are related by a quadratic
identification of variables $(u,v)\in\C^2$ with the $x$'s as
$x_1=u^2,x_2=v^2,x_3=uv$. To compare results of $x$'s to $(u,v)$ variables,
noting the $t$-charge assignment, a $t$ is to be replaced with a $t^2$ 
in the formulas for partition function \eq{pf:formula} as well as the 
Hilbert series. 
The comparison is, however, valid only in a local coordinate chart. Certain
monomials which survive the orbifold action in terms of $u,v$ variables are
absent in the description in terms of $x$'s. These correspond to missing
states in the latter description. For example, there are four monomials 
$u\pa u, u\pa v, v\pa u, v\pa v$ with grade $q t^2$ which survive the
orbifold projection. Only three of them appear in terms of $x$'s as 
$\pa x_1,\pa x_2$ and $\pa x_3 = u\pa v+v\pa u$. The combination 
$u\pa v-v\pa u $ is absent. Inclusion of this combination calls for
extending the set of regular functions on the variety $x_1x_2-x_3^2=0$ by 
rational functions of $\gamma$'s that is, $x$'s 
and their derivatives \cite{Grassi:2007va}.
This is achieved by including the blow up modes in the Hilbert series with
\eq{hbl}. While this mends the partition function at this level, states at
higher grades still remain missing. This may be ascribed to the fact that a
resolution of singularity by blowing up a point repairs the variety
$\mathcal{M}$ up to its tangent space, $\mathcal{M}_1$ in general. 
Thus we do not expect the partition function obtained without resorting to
the parametric representation to completely match \eq{zorb}. 
\subsection{Implementing Field-Antifield Symmetry}
Field-antifield symmetry possessed by the 
partition function takes the grade $t$ to its inverse.
The field-antifield symmetry can be imposed on the partition function
according to \eq{pf:g}. 
Taking $Z_{\gamma}= \H_{x_1x_2-x_3^2}$ the separation into
massless and massive modes is obvious. 
The zero mode part $Z_0(t)=(1-t^2)/(1-t)^3$ transforms to 
\begin{equation}
Z_0(t) = t^{-1} \, Z_0(1/t).
\end{equation}
Treating the $\beta$-$\gamma$ system as the ghost system of an 
appropriate string theory, the index $-1$ ($-2$ if compared to 
\eq{zorb}) of $t$ corresponds to $\gamma$-charge anomaly
\cite{Belavin:1984vu}. It indicates that
an appropriate number of antifields have to be introduced to 
define a consistent inner product on the full Hilbert space. 
At higher mass levels this is expected to be a symmetry, indicating that
all the physical states appear in field-antifield pairs. 
By \eq{gsplit},
\eq{pf:g} and \eq{hilb2:prod} this yields the total partition function
\begin{equation}
\label{pf:faf}
\hat{Z}(q,t)  = 
\frac{(1-t^2)}{(1-t)^3}
\prod\limits_{n=1}^{\infty}\frac{(1-q^nt^2)}{(1-q^nt)^3}
\frac{(1-q^n/t^2)}{(1-q^n/t)^3}.
\end{equation} 
Expanding in series in $q$ and $t$ this yields
\begin{equation}
\label{pf:kozul}
\begin{split}
\hat{Z}(q,t^2)  &= 
(1+3 t^2+5 t^4+7 t^6+9 t^8+11 t^{10}+\cdots)\\
&\;\; +q
(-{1}/{t^4}+4+11 t^2+20 t^4+28 t^6+36 t^8+44
   t^{10}+\cdots)\\
&\;\; +q^2
(-{3}/{t^6}-{4}/{t^4}+14+38 t^2+67 t^4+98 t^6+126
   t^8+154 t^{10}+\cdots)\\
&\;\;+q^3
   (-{5}/{t^8}-{11}/{t^6}-{14}/{t^4}+40+106 t^2+189
   t^4+275 t^6+360 t^8\\
&\qquad\qquad\qquad\qquad \qquad\qquad  +440 t^{10}+\cdots)\\
&\;\;+q^4
   (-{7}/{t^{10}}-{20}/{t^8}-{38}/{t^6}-{40}/{t^4}+
   105+275 t^2+487 t^4+715 t^6\\
&\qquad\qquad\qquad\qquad \qquad\qquad  +938 t^8+1155t^{10}+\cdots)\\
&\;\;+q^5
   (-{9}/{t^{12}}-{28}/{t^{10}}-{67}/{t^8}-{106}/{t^6}
-{105}/{t^4}+252+651 t^2+1154 t^4\\
&\qquad\qquad\qquad\qquad \qquad\qquad  +1697 t^6+2240 t^8+2763
   t^{10}+\cdots)+\mathcal{O}\left(q^6\right),
\end{split} 
\end{equation} 
where we have used the grade $t^2$ to compare with \eq{zorb}.
This exhibits missing states for terms with lower $t$-grades
for every power of $q$. Also, the series contains negative terms in 
the partition function which are difficult to account for. 
Adding the blow up modes \eq{hbl} exacerbates the mismatch, 
as noted earlier \cite{nek1}. Further examples of  partition function of beta-gamma systems,
such as, a system with constraint 
$\gamma^2=0$ and also the conifold, may be evaluated in this way matching
previous results \cite{gp}. 
\subsection{Implementing Gauge Invariance}
\revision{
Let us now work out the partition function using the gauge invariance of the
action, as explained in section \ref{sec:curv}.
In this method one first counts all monomials 
arising from $\gamma$'s and $\beta$'s 
separately as the Hilbert of series of the respective arc spaces of the
constraints \eq{constr:general} and \eq{constr:beta}, respectively. 
The partition function is then obtained by extracting the set of
gauge invariant monomials from \eq{pf:formula}.  
Let us illustrate this with examples, which will also justify the formula
\eq{pf:formula}. We shall use a generic symbol $\beta$ and $\gamma$ without
the indices for this purpose. 
\begin{example}
Let us compute the coefficient of $q^2/t$ in the partition function.
This grade is contributed by monomials of the generic form 
$\beta^2 \gamma$. The number of such combinations is obtained from the
coefficient of $q^2/t^2$ in \eq{ser:beta} and that of $q^0t$ in
\eq{ser:gamma}. The total number of such monomials respecting constraints
\eq{constr:general}  and \eq{constr:beta} is thus $5\times 3=15$.
Under the gauge transformation \eq{gagtr:general} a monomial of the form
$\beta^2\gamma$ goes to one of the form
$\beta\gamma^2$ changing the grade from $q^2/t$ to
$qt$, the change being a factor of $q/t^2$.
In order to count the ones that vanish modulo \eq{constr:general} we note
that the constraint is to be imposed on the portion of $\beta\gamma^2$ 
that is quadratic  in the $\gamma$'s. Now, $Z_{\gamma}$ counts the monomials
which \emph{do not} vanish modulo the constraint. Thus the number of
non-vanishing monomials of the form $\gamma^2$ is given by the coefficient of
$t^2$ in \eq{ser:gamma}, which is $5$. Multiplying with the $3$
$\beta$'s this gives the number of monomials  of the form $\beta\gamma^2$
that survive the constraint as $15$. Thus,
\begin{equation*} 
\begin{split}
\text{contribution of monomials of the form}\ \beta^2\gamma 
&=15q^2/t\\
\text{contribution of non-vanishing monomials of the form}\ \beta\gamma^2 
&=15qt
\end{split}
\end{equation*} 
As indicated in \eq{pf:formula},
the subtraction of the second term to obtain the number of vanishing ones
is effected in the partition function by multiplying the
it with $q/t^2$ compensating for the change of grade due to the
gauge transformation. We conclude that the number of
gauge invariant monomials is thus zero. The partition function does not have
a $q^2/t$ term in its series expansion.
\end{example}
\begin{example}
As the second example
let us consider the coefficient of the $q^2 t$ term in the partition
function. These arise from three types of monomials, \viz
$m_1=\beta^2\gamma^3$,
$m_2=\beta\gamma\pa\gamma$ and $m_3=\gamma^2\pa\beta$.
Number of $\beta^2$ is counted as $5$ from the coefficient of $q^2/t^2$ in
\eq{ser:beta}, while that of $\gamma^3$ is $7$ from the coefficient of the
$q^0t^3$ term of \eq{ser:gamma}, leading to $[m_1]=5\times7=35$ monomials of
type $m_1$. Under \eq{gagtr:general} these go over to monomials of the type
$\beta\gamma^4$. The number of non-vanishing monomials of this form,
corresponding to the gauge non-invariant combinations,
is counted as $3\times 9=27$ from the
coefficient of $q/t$ in \eq{ser:beta} and the coefficient of $t^4$ in
\eq{ser:gamma}.
Similarly, the number of monomials of type $m_2$ is obtained as
$[m_2]=3\times
8=24$ from the coefficient of $q/t$ in \eq{ser:beta} and that of $q t^2$ in
\eq{ser:gamma}. These give rise to monomials of the form
$\gamma^2\pa\gamma$. Non-vanishing combinations of this form, the gauge
non-invariant ones, are counted as the coefficient of $qt^3$ in
\eq{ser:gamma} to be $12$. Finally, the number of monomials of type $m_3$ is
$3\times 5=15$, obtained from the coefficient of $q^2/t$ in \eq{hilb2:beta}
and that of $t^2$ in \eq{ser:gamma}. Under \eq{gagtr:general} these go over
to monomials of the form $\gamma^2\pa\gamma$, which have been considered
above. Considering all these, the number of gauge invariant monomials are
\[
35+24+15-27-12=35.
\]
Adding the three $\pa^2\gamma$'s, namely, $\pa^2 x_1$, $\pa^2 x_2$ and $\pa^2
x_3$,  which contribute to this order as also seen
from the coefficient of $q^2t$ in \eq{ser:gamma}, the
coefficient of $q^2t$ in the partition function is $38$.
\end{example}
\begin{example}
\label{ex:neg}
For certain grades we end up with an over-determined system, however,
yielding negative coefficients. For example, monomials with grade 
$q^3/t^2$ arise from the $7\times 3=21$ monomials of the form
$\beta^3\gamma$, as seen from the coefficients of $q^3/t^3$ in \eq{ser:beta}
and that of $q^0t$ in \eq{ser:gamma}. Under the gauge transformation
\eq{gagtr:general} these produce terms of the form $\beta^2\gamma^2$. The
number of such monomials that survive modulo the constraint is $5\times 5$ as
seen from the coefficients of $q^2/t^2$ in \eq{ser:beta} and that of $q^0t^2$
in \eq{ser:gamma}. The difference $21-25=-4$ signifies that there is no gauge
invariant combination of the form $\beta^3\gamma$. 
\end{example}
This method thus provides a rationale for ignoring  
terms with negative coefficients in the partition function.

Generalizing these examples, we obtain \eq{pf:formula}. A monomial of the
form $\pa^{a}\beta^b\pa^c\gamma^d$ with grade $q^{a+b+c}t^{d-b}$ 
is counted from $Z_{\gamma}Z_{\beta}$ by multiplying the 
coefficients of $q^{a+b}/t^b$ from \eq{ser:beta} and
that of $q^c t^d$ from \eq{ser:gamma}. 
Under the transformation \eq{gagtr:general} such a monomial
reduces to $\pa^{a}\beta^{b-1}\pa^c\gamma^{d+1}$, with grade
$q^{a+b+c-1}t^{d-b+2}$. The corresponding partition function is then given
by $Z_{\beta}Z_{\gamma}q/t^2$. But the latter counts non-vanishing monomials
modulo the constraints. The gauge invariant ones are the vanishing ones. We
thus need to subtract them from $Z_{\gamma}Z_{\beta}$. However, monomials
formed from $\beta$'s alone can not be gauge invariant as each $\beta$
produces a single $\gamma$ under the gauge transformation \eq{gagtr:general},
while the constraint \eq{constr:general} is quadratic in $\gamma$'s. 
Moreover, a gauge transformation of a monomial of the form 
$\pa^{a}\beta^b\pa^c\gamma^d$ can not produce terms that are independent of
$\gamma$'s or linear in them. This explains the last two terms of
\eq{pf:formula}. 
}

Let us now present the complete formula.
Using the expression \eq{hilb2:prod} for $Z_{\gamma}$ and \eq{hilb2:beta}
for $Z_{\beta}$ in \eq{pf:formula}
we obtain the partition function as
\begin{equation}
\label{pf:unres}
\begin{split}
\tilde{Z}(q,t^2)
&=(1+3 t^2+5 t^4+7 t^6+9 t^8+11 t^{10}+\cdots)\\
&\;\; +q
   (4+11 t^2+20 t^4+28 t^6+36 t^8+44
   t^{10}+\cdots)\\
&\;\; +q^2(14+38 t^2+67 t^4+98
   t^6+126 t^8+154 t^{10}+\cdots)\\
&\;\; +q^3
   (-\frac{4}{t^4}+40+106 t^2+189 t^4+275 t^6+360 t^8+440
   t^{10}+\cdots)\\
&\;\;+q^4
   (-\frac{8}{t^6}-\frac{27}{t^4}+105+275 t^2+487 t^4+715 t^6+938
   t^8+1155 t^{10}+\cdots)\\
&\;\;+q^5
   (-\frac{12}{t^8}-\frac{49}{t^6}-\frac{86}{t^4}+252+651 t^2+1154
   t^4+1697 t^6\\
&\qquad\qquad\qquad\qquad\qquad\qquad +2240 t^8+2763
   t^{10}+\cdots)+\mathcal{O}(q^6),
\end{split}
\end{equation} 
which matches with \eq{pf:kozul} in the positive terms. 
Let us note that  the expression is written for $\tilde{Z}(q,t^2)$ rather
than $\tilde{Z}(q,t)$ to compare with \eq{pf:kozul}, which correspond to a
different assignment of charges. 
While this formula too is
plagued by the presence of negative terms, as discussed in 
Example~\ref{ex:neg} above, the negative terms may be ignored.

Incorporating certain rational functions by blowing up the
singularity, discussed before, ameliorates the results to a certain
extent.
Instead of  \eq{hilb2:prod} if we add the contribution of the exceptional
divisor of the blow up to the Hilbert series \eq{hbl} to set
\begin{equation} 
Z_{\gamma}=\H_{x_1x_2-x_3^2}+qt\ \H_{Bl_0(x_1x_2-x_3^2)},
\end{equation} 
where the factor $qt^2$ accounts for the unit  codimension of the 
exceptional divisors, 
then using this expressions with
\eq{hilb2:beta} and \eq{pf:formula} the corrected partition function reads
\begin{equation}
\label{a1:final}
\begin{split}
Z(q,t^2)&=
(1+3 t^2+5 t^4+7 t^6+9 t^8+11 t^{10}+\cdots)\\
&\;\;+q
   (4+12 t^2+20 t^4+28 t^6+36 t^8+44
   t^{10}+\cdots)\\
&\;\;+q^2 (17+38 t^2+70 t^4+98
   t^6+126 t^8+154 t^{10}+\cdots)\\
&\;\;+q^3
   (-\frac{4}{t^4}+\frac{5}{t^2}+40+115 t^2+189 t^4+280 t^6+360
   t^8+440 t^{10}+\cdots)\\
&\;\;+q^4
   (-\frac{8}{t^6}-\frac{20}{t^4}-\frac{1}{t^2}+123+279 t^2+502
   t^4+715 t^6+945 t^8+1155 t^{10}+\cdots)\\
&\;\;+q^5
   (-\frac{12}{t^8}-\frac{40}{t^6}-\frac{89}{t^4}+\frac{26}{t^2}+26
   4+685 t^2+1162 t^4+1718 t^6+2240 t^8\\
& \qquad \qquad \qquad \qquad \qquad +2772 t^{10}+\cdots)+\mathcal{O}(q^6).
\end{split} 
\end{equation} 
The partition function matches \eq{zorb} up to the first mass level
completely, which is expected since the first mass level corresponds to the
tangent space of the variety, which is repaired by a blow up.
\revision{
As discussed above, the negative terms correspond to a over-determined system
and need be ignored yielding the partition function for the blown up
rational double point
\begin{equation}
\label{a1:finalissimo}
\begin{split}
Z(q,t^2)&=
(1+3 t^2+5 t^4+7 t^6+9 t^8+11 t^{10}+\cdots)\\
&\;\;+q
   (4+12 t^2+20 t^4+28 t^6+36 t^8+44
   t^{10}+\cdots)\\
&\;\;+q^2 (17+38 t^2+70 t^4+98
   t^6+126 t^8+154 t^{10}+\cdots)\\
&\;\;+q^3
   (\frac{5}{t^2}+40+115 t^2+189 t^4+280 t^6+360
   t^8+440 t^{10}+\cdots)\\
&\;\;+q^4
   (123+279 t^2+502
   t^4+715 t^6+945 t^8+1155 t^{10}+\cdots)\\
&\;\;+q^5
   (\frac{26}{t^2}+264+685 t^2+1162 t^4+1718 t^6+2240 t^8\\
& \qquad \qquad \qquad \qquad \qquad +2772 t^{10}+\cdots)+\mathcal{O}(q^6).
\end{split} 
\end{equation}
}
Terms with sufficiently high order of 
$t^2$ match as well for all powers of $q$, indicating 
that the number of missing states are finite at each mass level.
As discussed in the examples above, the series can be verified at low orders in $q$ and $t^2$
by explicitly constructing all possible combinations of $\beta$'s and $x$'s with arbitrary coefficients and 
discarding the gauge non-invariant monomials.
\section{Partition function of Pure spinors}
\label{sec:ps}
In this section we present the results for the case of pure spinors. 
The action for the pure spinor system is
\begin{equation}
\label{ps:action}
S =  \frac{1}{2\pi}\int\omega^T\bar\pa\lambda, 
\end{equation} 
where 
$\lambda = \big(x_1,x_2,\cdots,x_{16} \big)^{T}$ 
is a sixteen dimensional complex vector subject to the constraint 
\begin{equation}
\label{ps:constraint}
{\lambda}\gamma^{\mu}\lambda=0 
\end{equation} 
and $\omega$ denotes its conjugate.
Here $T$ denotes matrix transpose and $\gamma^{\mu}$ denotes 
the ten-dimensional gamma matrices with
$\mu=0,1,\cdots,9$. 
The action possesses the classical gauge symmetry
\begin{equation} 
\delta_{\epsilon}\omega = 
\epsilon^{\mu}\gamma_{\mu}\lambda.
\end{equation} 
\revision{
We write down the partition function of  the pure spinor system using the
field-antifield symmetry \eq{pf:g}.
To  obtain $Z_{\gamma}$ we need to evaluate
the Hilbert series of the arc space of \eq{ps:constraint}.
The arc space is obtained by substituting the expansion \eq{pui} for the
sixteen complex coordinates in \eq{ps:constraint} leading to constraints
among the variables $x_i^{(j)}$. Since there are ten gamma matrices, there
are ten equations for every power of $\xi$, giving rise to the arc-space of
\eq{ps:constraint}. These are not algebraically independent, however. The
Hilbert series of the arc space counts the monomials in the variables
$x_i^{(j)}$ modulo the constraints defining the arc space.
Since there are more than
one equations at every mass level, the computation of Hilbert series is more
complicated than the previous case requiring the Gr\"obner basis of the ideal
generated by the constraints at each level, that is for each power of $q$.
We resort to {\tt Macaulay2} to compute the Hilbert series. The
first jet  scheme of the variety \eq{ps:constraint} is obtained by writing 
\begin{equation}
\label{lambda:pui}
\lambda = \lambda^{(0)}+\lambda^{(1)}\xi:=
\big(x_1^{(0)}+x_1^{(1)}\xi,x_2^{(0)}+x_2^{(1)}\xi,
\cdots,x_{16}^{(0)}+x_{16}^{(1)}\xi \big)^{T} 
\end{equation} 
and substituting in \eq{ps:constraint}. The twenty resulting equations,
namely,
\begin{gather}
{\lambda}^{(0)}\gamma^{\mu}\lambda^{(0)}=0 \\
{\lambda}^{(0)}\gamma^{\mu}\lambda^{(1)} 
+{\lambda}^{(1)}\gamma^{\mu}\lambda^{(0)}=0 
\end{gather}
define the first jet scheme. 
The Hilbert series is obtained from the set of equations with grades $t$ for
$\lambda^{(0)}$ and $qt$ for $\lambda^{(1)}$. Computing the Hilbert series in
{\tt Macaulay2} yields 
\begin{equation}
\label{lev1}
\begin{split}
\big(1 &+5 t-10 q t^{2}+5 t^{2}-34 q t^{3}+t^{3}+q^{3} t^{4}+45
q^{2} t^{4}-16 q t^{4}\\
&-11 q^{3} t^{5}+65 q^{2} t^{5}-65 q^{3} t^{6} +11 q^{2}
t^{6}+16 q^{4} t^{7}-45 q^{3} t^{7}-q^{2} t^{7}-q^{5} t^{8}\\
&+34 q^{4} t^{8}-5
q^{5} t^{9}+10 q^{4} t^{9}
-5 q^{5} t^{10}-q^{5} t^{11}\big)/({1-t})^{11} ({1-q t})^{16}.
\end{split}
\end{equation}
This result is correct to order $q$ only since we truncated the 
series \eq{lambda:pui} at $\lambda^{(1)}$. 
}
Expanded in powers of $q$ and retaining term up to order $q$ we obtain
\begin{equation}
{\mathcal H}_{{\lambda}\gamma^{\mu}\lambda}
=\frac{t^3+5 t^2+5 t+1}{(1-t)^{11}}+q\frac{2\left(23 t^3+35 t^2+8
   t\right)}{(1-t)^{11}}+O\left(q^2\right),
\end{equation}
In order to implement the field-antifield
symmetry according to \eq{pf:g} the Hilbert series is taken to be the
contribution of $\gamma$'s to the partition function. We write it as
\begin{equation} 
Z_{\gamma}(q,t)=
{\mathcal H}_{{\lambda}\gamma^{\mu}\lambda} =
\frac{1+5t+5t^2+t^3}{(1-t)^{11}} \tilde{Z}(q,t)
\end{equation} 
by pulling out the factor $Z_0(t)=(1+5t+5t^2+t^3)/(1-t)^{11}$. Then the
partition function of the beta-gamma
system \eq{ps:action} with the pure spinor
constraint \eq{ps:constraint} is given by \eq{pf:g} as
\begin{equation} 
Z_{PS}(q,t) = \frac{1+5t+5t^2+t^3}{(1-t)^{11}} 
\tilde{Z}(q,t) \tilde{Z} (q,1/t). 
\end{equation} 
Expanded in powers of $q$ this yields
\begin{equation} 
Z_{PS}(q,t) = \frac{t^3+5 t^2+5 t+1}{(1-t)^{11}}+q\frac{2(t+1) \left(23 t^2+20
   t+23\right)}{(1-t)^{11}}+\mathcal{O}(q^2).
\end{equation} 
This matches with the expression obtained earlier \cite{aabn} up to the first
mass level.
\section{Conclusion}
\label{sec:concl}
We obtain the partition function of beta-gamma systems with
algebraic constraints on the fields $\gamma$. We showed that the partition
function of a beta-gamma system can be evaluated by identifying the
contribution from the $\gamma$'s as the Hilbert series of arc spaces of the
algebraic variety given by the constraint. 
Two examples are worked out
explicitly. In the first we consider the $A_1$ surface singularity given by
a quadratic constraint in three $\gamma$'s. The partition function evaluated
using the constraint without solving it with a parametric representation is
expected to be different from that obtained using its description as an
orbifold, dealt with in an earlier publication \cite{ours}. 
We demonstrate two different 
ways of computing the partition function in this case. The first one
implements the so-called field-antifield symmetry in a multiplicative
fashion. This, however, gives rise to terms with negative coefficients in the
partition function, which can not be accounted for as the partition 
is the generating function of degeneracy of operators. 
We show that the partition function can be obtained alternatively
as the generating function of monomials invariant under
the classical gauge symmetry of the action modulo the constraint. 
This is implemented by subtracting the number of monomials not vanishing
under the gauge transformation from the totality of monomials. 
Hence the terms with negative coefficients signify an over-determined system
and may thus be omitted. The positive terms of both the 
expressions, on the other hand, match. 
Moreover, using the description of the 
blow up of the codimension two singularity in terms of arc spaces it is shown
that the partition function matches with the orbifold partition function up
to the first mass level, as the blow up repairs the tangent space of the
orbifold. An advantage of the algorithm 
presented here lies in
the fact that it can be straightforwardly extended to the case where the 
constraints are not reducible, such as pure spinor system. 
Moreover, this gives the partition function a geometric significance.
A computation using Poisson brackets confirms \eq{a1:finalissimo} at the lowest
grades.

We also obtain the partition function of the pure spinor system using the
Hilbert series of the arc space of the pure spinor constraint looked upon as
a variety embedded in the sixteen-dimensional complex affine space. This
requires the computation of Gr\"obner bases in the polynomial rings involved.
We used \verb|Macaulay2| to obtain the Hilbert series which, though
straightforward  as an algorithm, is
extremely memory-intensive and we are restricted here to 
the first mass level. The code is appended below. However, we show that the
computation up to this level implementing the field-antifield symmetry in a
product formula matches with previously known results \cite{aabn}. The
computation of gauge invariant monomials is more complicated since the
resolution in terms of arc spaces is not known in addition to the
variety of gauge invariants appearing at higher mass levels. \\

\noindent{\bf{Acknowledgments:}} C.B. thanks IIT Bhubaneswar for seed project
SP-0038. We thank the anonymous referee for useful suggestions.

\clearpage
\section*{
Appendix: the na\"ive {\tt Macaulay2} code used for purespinors}
{\tiny 
\begin{verbatim}
baseRing = ZZ;

makeVars = (N,K) -> flatten toList apply(0..N, I -> flatten toList
apply(1..K, a -> x_[I,a]));

-- X_Ia:    I := mass level
--          a := index of fields, 1 ... 16


lstDeg = (N,K) -> flatten toList apply(0..N, I -> toList apply(1..K, a ->
{I,1}));

-- degrees corresponding to mass level change
-- all fields (in the a index) are equal degree

-- R = baseRing[makeVars(1,16),Degrees => lstDeg(1,16)];
                                   -- CHANGE N of (N,K) for each mass level




unit = matrix{{1,0},{0,1}}
tau1 = matrix{{0,1},{1,0}}
ep = matrix{{0,1},{-1,0}}    -- this is (i tau2)
tau3 = matrix{{1,0},{0,-1}}

g1 =  ep ** ep ** ep
g2 =  unit ** tau1 ** ep
g3 =  unit ** tau3 ** ep
g4 =  tau1 ** ep ** unit
g5 =  tau3 ** ep ** unit
g6 =  ep ** unit ** tau1
g7 =  ep ** unit ** tau3
g8 =  unit ** unit ** unit

zer = id_(R^8) * 0

G1 = matrix{{zer,g1},{transpose(g1),zer}}
G2 = matrix{{zer,g2},{transpose(g2),zer}}
G3 = matrix{{zer,g3},{transpose(g3),zer}}
G4 = matrix{{zer,g4},{transpose(g4),zer}}
G5 = matrix{{zer,g5},{transpose(g5),zer}}
G6 = matrix{{zer,g6},{transpose(g6),zer}}
G7 = matrix{{zer,g7},{transpose(g7),zer}}
G8 = matrix{{zer,g8},{transpose(g7),zer}}
G9 = id_(R^16)
G0 = matrix{{id_(R^8),zer},{zer,-id_(R^8)}}

-- MASS LEVEL 0

m0 = matrix{toList apply(1..16, a -> x_[0,a])}
m0t = transpose(m0)

m0Ideal1 = m0*G1*m0t
m0Ideal2 = m0*G2*m0t
m0Ideal3 = m0*G3*m0t
m0Ideal4 = m0*G4*m0t
m0Ideal5 = m0*G5*m0t
m0Ideal6 = m0*G6*m0t
m0Ideal7 = m0*G7*m0t
m0Ideal8 = m0*G8*m0t
m0Ideal9 = m0*G9*m0t
m0Ideal0 = m0*G0*m0t

-- MASS LEVEL 1

m1 = matrix{toList apply(1..16, a -> x_[1,a])}
m1t = transpose(m1)

m1Ideal1 = m1*G1*m0t + m0*G1*m1t
m1Ideal2 = m1*G2*m0t + m0*G2*m1t
m1Ideal3 = m1*G3*m0t + m0*G3*m1t
m1Ideal4 = m1*G4*m0t + m0*G4*m1t
m1Ideal5 = m1*G5*m0t + m0*G5*m1t
m1Ideal6 = m1*G6*m0t + m0*G6*m1t
m1Ideal7 = m1*G7*m0t + m0*G7*m1t
m1Ideal8 = m1*G8*m0t + m0*G8*m1t
m1Ideal9 = m1*G9*m0t + m0*G9*m1t
m1Ideal0 = m1*G0*m0t + m0*G0*m1t


arc1 = ideal(
m0Ideal1,
m0Ideal2,
m0Ideal3,
m0Ideal4,
m0Ideal5,
m0Ideal6,
m0Ideal7,
m0Ideal8,
m0Ideal9,
m0Ideal0,
m1Ideal1,
m1Ideal2,
m1Ideal3,
m1Ideal4,
m1Ideal5,
m1Ideal6,
m1Ideal7,
m1Ideal8,
m1Ideal9,
m1Ideal0
);

hf = hilbertSeries arc1
reduceHilbert hf
\end{verbatim}
}


\begin{thebibliography}{99}
\bibitem{wit1} 
E.~Witten,
``Two-dimensional models with (0,2) supersymmetry: Perturbative aspects,''
Adv.\ Theor.\ Math.\ Phys.\  {\bf 11} (2007) [hep-th/0504078].
\bibitem{nek1}
N.~A.~Nekrasov,
``Lectures on curved beta-gamma systems, pure spinors, and anomalies,''
[hep-th/0511008].
\bibitem{Tan:2006qt} 
  M.~-C.~Tan,
  ``Two-dimensional twisted sigma models and the theory of chiral differential operators,''
  Adv.\ Theor.\ Math.\ Phys.\  {\bf 10}, 759 (2006)
  [hep-th/0604179].


\bibitem{gp}
P.~A.~Grassi and G.~Policastro,
``Curved beta-gamma systems and quantum Koszul resolution,''
[hep-th/0602153].
\bibitem{Grassi:2007va} 
  P.~A.~Grassi, G.~Policastro and E.~Scheidegger,
  ``Partition Functions, Localization, and the Chiral de Rham complex,''
  [hep-th/0702044].
\bibitem{aisaka} 
Y.~Aisaka and E.~A.~Arroyo,
``Hilbert space of curved beta gamma systems on quadric cones,''
JHEP {\bf 0808}, 052 (2008) [arXiv:0806.0586 [hep-th]].

\bibitem{ours} 
  C.~Bhamidipati and K.~Ray,
  ``Partition function of beta-gamma system on orbifolds,''
  JHEP {\bf 1311}, 152 (2013)
  [arXiv:1308.5117 [hep-th]].

\bibitem{berkovits} N. Berkovits, ``Super Poincare covariant quantization of
the superstring,''
  JHEP {\bf 0004}, 018 (2000),  [hep-th/0001035].


\bibitem{Berkovits:2005hy} 
  N.~Berkovits and N.~Nekrasov,
  ``The Character of pure spinors,''
  Lett.\ Math.\ Phys.\  {\bf 74}, 75 (2005)
  [hep-th/0503075].

\bibitem{morales} 
P.~A.~Grassi and J.~F.~Morales Morera,
``Partition functions of pure spinors,''
Nucl.\ Phys.\ B {\bf 751}, 53 (2006)
[hep-th/0510215].


\bibitem{aabn} 
Y.~Aisaka, E.~A.~Arroyo, N.~Berkovits and N.~Nekrasov,
``Pure Spinor Partition Function and the Massive Superstring Spectrum,''
JHEP {\bf 0808}, 050 (2008) [arXiv:0806.0584 [hep-th]].


\bibitem{arroyo} 
E.~Aldo Arroyo,
``Pure Spinor Partition Function Using Pade Approximants,''
JHEP {\bf 0807}, 081 (2008)
[arXiv:0806.0643 [hep-th]].


\bibitem{m2}
Grayson, Daniel R. and Stillman, Michael E., 
`` Macaulay2, a software system for research in algebraic geometry''.
Available at {\tt http://www.math.uiuc.edu/Macaulay2/}


\bibitem{Guruswamy} 
  S.~Guruswamy and A.~W.~W.~Ludwig,
  ``Relating $c < 0$ and $c >$ 0 conformal field theories,''
  Nucl.\ Phys.\ B {\bf 519}, 661 (1998)
  [hep-th/9612172].
\bibitem{Lesage} 
  F.~Lesage, P.~Mathieu, J.~Rasmussen and H.~Saleur,
  ``The $\hat{su}(2)_{-1/2}$ WZW model and the $\beta\gamma$ system,''
  Nucl.\ Phys.\ B {\bf 647}, 363 (2002)
  [hep-th/0207201].
\bibitem{nash}
J.\ F.\ Nash,\ Jr, ``Arc structure of singularities",
Duke Math. J. {\bf 81}.31.(1995).
\bibitem{rr:arc}
C.\ Bruschek, H.\ Mourtada, J.\ Schepers,
``Arc Spaces and Rogers-Ramanujan Identities",
[arXiv:1101.4950 (math)].
\bibitem{stanley}
R.~P.~Stanley, ``
Invariants of finite groups and their applications to combinatorics",
Bull. Amer. Math. Soc. {\bf 1} (1979), 475.
\bibitem{reguera}
A.\ J.\ Reguera, ``Families of arcs on rational surface singularities,''
Manuscripta Mathematica {\bf 88},\ 321 (1995).
\bibitem{Belavin:1984vu}
  A.~A.~Belavin, A.~M.~Polyakov and A.~B.~Zamolodchikov,
  ``Infinite conformal symmetry in two-dimensional quantum field theory,''
  Nucl.\ Phys.\  B {\bf 241} (1984) 333.
\end{thebibliography}
\end{document}